\documentclass[%
reprint,
 amsmath,amssymb,nofootinbib,
aps,
]{revtex4-2}

\usepackage{graphicx}
\usepackage{dcolumn}
\usepackage{bm}
\usepackage[normalem]{ulem}
\usepackage{hyperref}

\usepackage{physics}
\usepackage{xcolor}

\begin{document}


\title{Can the diffeomorphism and Gauss constraints be holonomy corrected\\ 
in the deformed algebra approach to modified gravity?}

\author{Jamy-Jayme Thézier}
\email{jamyjayme.thezier@gmail.com}
\author{Aurélien Barrau}
\author{Killian Martineau}
\author{Maxime De Sousa}

\affiliation{
 Laboratoire de Physique Subatomique et de Cosmologie, Univ. Grenoble Alpes, CNRS/IN2P3\\ 53 avenue des Martyrs, 38026 Grenoble Cedex, France
}

\date{\today}

\begin{abstract}
Deforming the algebra of constraint is a well-known approach to  effective loop quantum cosmology. More generally, it is a consistent way to modify gravity from the Hamiltonian perspective. In this framework, the Hamiltonian (scalar) constraint is usually the only one to be holonomy corrected. As a heuristic hypothesis, we consider the possibility to also correct the diffeomorphism and Gauss constraints. It is shown that it is impossible to correct the diffeomorphism constraint without correcting the Gauss one, while maintaining a first-class algebra. However, if all constraints are corrected, the algebra can be closed. The resulting differential equations to be fulfilled by the corrections (of the background and of the perturbations) are derived.
\end{abstract}

\maketitle

\section{Introduction}


This article focuses on the deformation of the algebra of general relativity. This approach has been recently intensively considered as a way to effectively describe loop quantum cosmology \cite{Bojowald:2008jv, Wu:2012mh, Mielczarek:2011ph, Cailleteau:2011kr, Cailleteau:2012fy, Cailleteau:2013kqa, Barrau:2014maa, Bolliet:2015bka,BenAchour:2016leo, Martineau:2017tdx,Han:2017wmt, Barrau:2018gyz,Han:2018usc,Renevey:2021tmh,DeSousa:2022rep,Li:2023axl, DeSousa:2025cru}. In fact, it is much more general and can be viewed as a rigorous way to study modified gravity in the Hamiltonian setting. We intentionally keep this work as brief and concise as possible. As such, we do not go through the motivations and the framework behind. We refer any interested reader to \cite{DeSousa:2024icf} and references therein both for the full picture and for the details about the notations.\\ 

In the loop view, quantum geometry corrections are effectively taken into account by replacing the Sen-Ashtekar-Barbero connection by a generalized function that encodes the holonomy along a graph, the so-called holonomy correction. Following recent works -- in particular remarks made in \cite{Perez:2005fn,Vandersloot:2005kh,BenAchour:2016ajk,Amadei:2022zwp} -- 
we implement this correction into the classical algebra of general relativity, without specifying its explicit form {\it a priori}. Then, we ensure the consistency of the model by requiring the closure of the algebra of constraints -- that is, in making sure that each Poisson bracket between constraints is proportional to a constraint, or a combination of constraints: the evolution of a physical solution yields another physical solution. This approach actually extends far beyond its initial purpose and offers a powerful framework for studying modified gravity.

 To ensure the closure of the algebra of constraints, we introduce counterterms
 to cancel the anomalies induced in the algebra by the holonomy correction. These counterterms can also be regarded as additional terms introduced to construct the most general theory — under some natural hypotheses — that would incorporate the relevant corrections and would lead to general relativity in the classical limit. In this view, our results are not limited to the loop framework.

At the end of the day, one should obtain a first-class Poisson algebra and the subtle cancellation game between anomalies and counterterms could lead to restrictions on the allowed forms of the correction functions. This might have observable consequences as the Mukhanov-Sasaki equation for gauge-invariant perturbations is expected to be modified accordingly. Not to mention a possible disappearance of time \cite{Bojowald:2015gra,Barrau:2016sqp} around the quantum bounce \cite{Ashtekar:2011ni}.\\

Usually, the Hamiltonian constraint is the only one to receive corrections. Although there are reasons for this \cite{Bojowald:2012ux}, we investigate here, at the heuristic level, the possibility that the diffeomorphism and Gauss constraints may themselves receive corrections. We begin by focusing on the correction of the diffeomorphism constraint alone. We then proceed by considering the full set of corrections affecting all the constraints that define the theory. The differential equations to be fulfilled by the corrections are investigated in detail. We finally conclude with some remarks on possible future studies. 

\section{Correcting only the Hamiltonian and diffeomorphism constraints}\label{sec:correc diffeo}

\subsection{Expression of the constraints}

As in \cite{DeSousa:2024icf}, the Hamiltonian constraint $\mathbb{H}$ is corrected by two distinct functions $g(\mathfrak{c},\mathfrak{p})$ and $\Tilde{g}(\mathfrak{c},\mathfrak{p})$, which respectively replace the curvature $\mathfrak{c}$ at the background and perturbative levels. The variables $\mathfrak{p}$ and $\mathfrak{c}$, defined right after Eq. (\ref{eq:diffeo_density_corrected}), respectively describe the scale factor and the expansion rate.\\

In this work, in addition to the Hamiltonian constraint, the diffeomorphism constraint is also corrected by a function $h(\mathfrak{c},\mathfrak{p})$. Following the usual procedure, three new counterterms $\xi_i(\mathfrak{c},\mathfrak{p})$, $i=1,2,3$, are then introduced. The geometrical constraint reads :

\begin{equation}\label{eq:diffeo_constraint_def}
     \mathbb{D}_\mathfrak{g}\left[ N^a\right]=\kappa^{-1} \int d{\bf{x}}\delta N^a \mathcal{D}^{\mathfrak{g}(1)}_a,
 \end{equation}
 
 where
 
 \begin{align}
     \mathcal{D}^{\mathfrak{g}(1)}_a &=\mathfrak{p}(1+\xi_1)\partial_a \delta K_b^b-\mathfrak{p}(1+\xi_2)\partial_i \delta K^i_a \notag\\&- (h+\xi_3) \delta^j_a \partial_b \delta E^b_j,\label{eq:diffeo_density_corrected}
 \end{align}

with $N^a={\bf{N}}^a + \delta N^a$ the shift function, $E_i^a = \mathfrak{p} \delta_i^a + \delta E^a_i$ the densitized triads, and $K_a^i=\mathfrak{c} \delta^i_a + \delta K^i_a$ their canonically conjugate variables. Each of those quantities are perturbed around a homogeneous and isotropic background.\\

As usual, the matter sector of the diffeomorphism constraint is uncorrected -- as is the Gauss constraint for now. The expressions of each constraint are given in Appendix \ref{sec:appendix_expconstraints}.

\subsection{Cancellation of the anomalies}

Correcting the diffeomorphism constraint leads to a modification of the algebra of constraint and, in particular, to additional anomalies when compared to \cite{DeSousa:2024icf}. Those new anomalies are listed in Appendix \ref{sec:list anomalies juste diffeo}. To keep this article as short as possible, we do not make explicit the heavy calculations and we simply focus on the solutions. The approach consists of expressing all the counterterms as functions of the corrections, and then constraining the corrections if any anomalies remain.\\

The conditions $\mathcal{A}_2^{\left\{ \mathbb{H},\mathbb{D} \right\}}=0$ and $\mathcal{A}_7^{\left\{ \mathbb{H},\mathbb{D} \right\}}=0$ (\ref{anomalie_diffeo:H,D 2} and \ref{anomalie_diffeo:H,D 7}) imply $\xi_1=\xi_2$ as $\alpha_9$ and $\alpha_3$ cannot be constant so as to ensure a correct classical limit. The same limit considered for $\alpha_3$ and $\alpha_1$, together with $\mathcal{A}_5^{\left\{ \mathbb{H},\mathbb{H} \right\}}=0$ (\ref{anomalie_diffeo:H,H 5}), implies

\begin{equation}\label{eq_diffeo : xi_1 condition}
\xi_1 = 0 
\end{equation}

and, thus,

\begin{equation}\label{eq_diffeo : xi_2 condition}
    \xi_2=0.
\end{equation}

These counterterms are therefore ``not physical" or, if one sees them as mere corrections, not necessary.
Then, $\mathcal{A}_1^{\left\{ \mathbb{D},\mathbb{G} \right\}}=0$ (\ref{anomalie_diffeo:D,G 1}), leads, with Eq. (\ref{eq_diffeo : xi_2 condition}), to

\begin{equation}\label{eq_diffeo: h+xi_3 condition}
    \xi_3 = \mathfrak{c}-h.
\end{equation}

Since $h$ and $\xi_3$ appear only as $h+\xi_3=\mathfrak{c}$ in the anomalies, the system reduces to the one obtained without correcting the diffeomorphism constraint (see \cite{DeSousa:2024icf} for the end of the resolution). There is no other condition on $h$ and $\xi_3$ to ensure that $\mathcal{A}_1^{\left\{ \mathbb{D},\mathbb{G} \right\}}$ vanishes. 
As long as the counterterm $\xi_3$ compensates $h$ exactly, the algebra is first class and the theory remains consistent. This kind of trivial compensation -- which does not happen in the usual case where only the Hamiltonian constraint is corrected -- is equivalent to not correcting the diffeomorphism constraint.
Interestingly, had $\xi_3$ not been introduced initially, Eq. (\ref{eq_diffeo: h+xi_3 condition}) would impose $h=\mathfrak{c}$, but with $\xi_3$ included, it becomes possible for $h$ to differ from $\mathfrak{c}$.

It is important to emphasize that there are no additional degrees of freedom when compared to \cite{DeSousa:2024icf}. All the counterterms except $\alpha_9$ are fixed by the corrections.\\

With a holonomy-corrected Hamiltonian constraint, it is therefore not possible to correct the diffeomorphism constraint \textit{alone}\footnote{By "alone" we mean without correcting the Gauss constraint.} while maintaining a first-class algebra.


\section{Correcting all constraints}

\subsection{Expression of the constraints}

We now investigate what happens if all the constraints are holonomy corrected. In particular, Eq. (\ref{eq_diffeo: h+xi_3 condition}) links the diffeomorphism correction to the Gauss constraint, which makes the situation quite different if the latter is also corrected. This anomaly changes and the algebra of constraints can be closed differently.\\

Is is now assumed that the Gauss constraint is corrected with a new function $l(\mathfrak{c},\mathfrak{p})$. Associated  counterterms $\chi_1(\mathfrak{c},\mathfrak{p})$ and $\chi_2(\mathfrak{c},\mathfrak{p})$ are also introduced, so that the constraint reads,

\begin{equation}\label{eq_gauss: def gauss constraint}
    \mathbb{G}[\Lambda^i]=\kappa^{-1} \int d{\bf{x}} \; \delta \Lambda^i \mathcal{G}_i,
\end{equation}

with

\begin{equation}\label{eq_gauss: def density gauss constraint}
    \mathcal{G}_i = (\mathfrak{p}+\chi_1) \varepsilon_{ij}^a \delta K_a^j + (l+\chi_2) \varepsilon_{ia}^j \delta E^a_j,
\end{equation}

$\Lambda^i= {\bf{\Lambda}}^i + \delta \Lambda ^i$  being the perturbed Lagrange multipliers.

\subsection{Cancellation of the anomalies}

The new list of anomalies is given in Appendix \ref{sec:list anomalies tout corrigé}. This section focuses on the resolution of the associated system.\\

First, $\mathcal{A}_5^{\{ \mathbb{H},\mathbb{H} \}}=0$ (\ref{anomalie_gauss:H,H 5}) implies $\xi_1=0$. Combined with $\mathcal{A}_2^{\{ \mathbb{H},\mathbb{D} \}}=0$ and $\mathcal{A}_7^{\{ \mathbb{H},\mathbb{D} \}}=0$ (\ref{anomalie_gauss:H,D 2} and \ref{anomalie_gauss:H,D 7}), the classical limits of $\alpha_9$ and $\alpha_3$ imply

\begin{equation}
    \xi_1=\xi_2=0.
\end{equation}

This causes the anomalies $\mathcal{A}_8^{\{ \mathbb{H},\mathbb{D} \}}$ and $\mathcal{A}_9^{\{ \mathbb{H},\mathbb{D} \}}$ (\ref{anomalie_gauss:H,D 8} and \ref{anomalie_gauss:H,D 9}) to vanish immediately.\\

Considering the classical limits, the cancellation of $\mathcal{A}_{16}^{\{ \mathbb{H},\mathbb{D} \}}$ and $\mathcal{A}_{17}^{\{ \mathbb{H},\mathbb{D} \}}$ (\ref{anomalie_gauss:H,D 16} and \ref{anomalie_gauss:H,D 17}) gives

\begin{equation}\label{eq:beta4=beta8=0}
    \beta_8=\beta_4=0,
\end{equation}

leading to $\mathcal{A}_{9}^{\{ \mathbb{H},\mathbb{H} \}}=0$ (\ref{anomalie_gauss:H,H 9}). The cancellations of $\mathcal{A}_{10}^{\{ \mathbb{H},\mathbb{D} \}}$, $\mathcal{A}_{11}^{\{ \mathbb{H},\mathbb{D} \}}$ and $\mathcal{A}_{15}^{\{ \mathbb{H},\mathbb{D} \}}$ (\ref{anomalie_gauss:H,D 10}, \ref{anomalie_gauss:H,D 11} and \ref{anomalie_gauss:H,D 15}) imply, respectively,

\begin{align}
    &\beta_2=2\beta_1,\\
    &\beta_5=\beta_6 \label{eq:beta5=beta6},\\
    &\beta_{11}=\beta_{12}.
\end{align}

Adding $\mathcal{A}_{12}^{\{ \mathbb{H},\mathbb{D} \}}=0$ and $\mathcal{A}_{13}^{\{ \mathbb{H},\mathbb{D} \}}=0$ (\ref{anomalie_gauss:H,D 12} and \ref{anomalie_gauss:H,D 13}) gives $\beta_{10}=\beta_{12}=\beta_{11}$. Reinjecting this result in $\mathcal{A}_{13}^{\{ \mathbb{H},\mathbb{D} \}}=0$ and subtracting the latter from $\mathcal{A}_{12}^{\{ \mathbb{H},\mathbb{D} \}}=0$ leads to

\begin{equation}
    \beta_{10}=\beta_{11}=\beta_{12}=\partial_\mathfrak{c}(h+\xi_3)-1.
\end{equation}

Then, to be consistent with the classical limit of $\alpha_3$, $\mathcal{A}_{4}^{\{ \mathbb{H},\mathbb{H} \}}=0$ (\ref{anomalie_gauss:H,H 4}) implies 

\begin{equation}
    \alpha_4=\alpha_5.
\end{equation}

Canceling $\mathcal{A}_{6}^{\{ \mathbb{H},\mathbb{D} \}}$ (\ref{anomalie_gauss:H,D 6}) then gives 

\begin{equation}
    \alpha_7=\alpha_8.
\end{equation}

Since $\beta_4=0$ (\ref{eq:beta4=beta8=0}), $\mathcal{A}_{15}^{\{ \mathbb{H},\mathbb{H} \}}=0$ (\ref{anomalie_gauss:H,H 15}) implies

\begin{equation}\label{eq:relation alpha 1 alpha 4 alpha5}
    \partial_\mathfrak{c}(\Tilde{g}+\alpha_1) = \frac{1}{2} (2-\alpha_4 + 3 \alpha_5).
\end{equation}

Reinjecting this result in $\mathcal{A}_{14}^{\{ \mathbb{H},\mathbb{H} \}}=0$ (\ref{anomalie_gauss:H,H 14}), and making use of $\alpha_4=\alpha_5$, leads to $2\beta_2 (1+\alpha_4)=0$. Thus, recovering the correct classical limit for $\alpha_4$ imposes $\beta_2=0$. Since $2\beta_1=\beta_2$, this gives

\begin{equation}
    \beta_1=\beta_2=0.
\end{equation}

Considering $\mathcal{A}_{13}^{\{ \mathbb{H},\mathbb{H} \}}=0$ (\ref{anomalie_gauss:H,H 13}) leads to

\begin{equation}\label{eq:relat beta6 beta3}
    \beta_5=\beta_6=-\frac{\beta_3}{1+\beta_3},
\end{equation}

since $\beta_5=\beta_6$, from Eq. (\ref{eq:beta5=beta6}). The cancellation of $\mathcal{A}_{17}^{\{ \mathbb{H},\mathbb{H} \}}$ (\ref{anomalie_gauss:H,H 17}) and $\beta_1=0$ imply

\begin{equation}\label{eq:expression beta3}
    \beta_3=\frac{\Tilde{g}+\alpha_1}{g \partial_\mathfrak{c} g}-1.
\end{equation}

Inserting this expression with $\mathcal{A}_{7}^{\{ \mathbb{H},\mathbb{H} \}}=0$ (\ref{anomalie_gauss:H,H 7}), which implies $\partial_\mathfrak{c} \beta_3 = 0$, in $\mathcal{A}_{20}^{\{ \mathbb{H},\mathbb{H} \}}=0$ (\ref{anomalie_gauss:H,H 20}) leads to

\begin{equation}
    \partial_\mathfrak{p}\beta_3=0.
\end{equation}

This also relies on the fact that $g$ and $\partial_\mathfrak{c}g$ cannot vanish, to get the correct classical limit ($g\rightarrow \mathfrak{c}$). The term $\beta_3$ is then a constant since it depends neither on $\mathfrak{p}$ nor on $\mathfrak{c}$ ($\mathcal{A}_{7}^{\{ \mathbb{H},\mathbb{H} \}}=0$). As a consequence, the only way to get the correct classical limit is to have 

\begin{equation}
    \beta_3=0.
\end{equation}

Considering Eq. (\ref{eq:relat beta6 beta3}), this gives

\begin{equation}
    \beta_5=\beta_6 = 0.
\end{equation}

This cancellation applied to $\mathcal{A}_{14}^{\{ \mathbb{H},\mathbb{D} \}}$ (\ref{anomalie_gauss:H,D 14}) implies  $\beta_{10}=-\beta_9$, and, thus, 

\begin{equation}\label{eq:relation9 10 11 12 beta et correc diffeo}
    \beta_{10}=-\beta_9=\beta_{12}=\beta_{11} = \partial_\mathfrak{c}(h+\xi_3)-1.
\end{equation}

Equation (\ref{eq:expression beta3}) must hold with $\beta_3=0$ and implies

\begin{equation}\label{eq:expr alpha 1}
    \alpha_1=g \partial_\mathfrak{c}g - \Tilde{g}.
\end{equation}

The cancellation of $\mathcal{A}_{12}^{\{ \mathbb{H},\mathbb{H} \}}$ (\ref{anomalie_gauss:H,H 12}) leads to

\begin{equation}
    \beta_7=0.
\end{equation}

The only nonvanishing matter counterterms are, thus, $\beta_{13}$, which is still undetermined, and $\beta_{10}$, $\beta_9$, $\beta_{12}$, and $\beta_{11}$ which are related, by Eq. (\ref{eq:relation9 10 11 12 beta et correc diffeo}), to the diffeomorphism correction. Those latter counterterms would vanish if the diffeomorphism constraint was not corrected.\\

At this level, $\mathcal{A}_{1}^{\{ \mathbb{H},\mathbb{D} \}}=0$ (\ref{anomalie_gauss:H,D 1}) implies

\begin{equation}
    \alpha_2 = 3 g^2 - \Tilde{g}^2 -2g  (h+\xi_3)\partial_\mathfrak{c} g.
\end{equation}

Taking into account that $\alpha_4=\alpha_5$ in Eqs. (\ref{eq:relation alpha 1 alpha 4 alpha5}) and (\ref{eq:expr alpha 1}) gives

\begin{equation}\label{eq: expr alpha 4 et alpha 5}
    \alpha_4=\alpha_5= (\partial_\mathfrak{c}g)^2 + g \partial^2_\mathfrak{c} g -1.
\end{equation}

Canceling $\mathcal{A}_{3}^{\{ \mathbb{H},\mathbb{D} \}}$ (\ref{anomalie_gauss:H,D 3}) then implies

\begin{equation}
    \alpha_6 = 2 g \partial_\mathfrak{c}g - \Tilde{g} - (h+\xi_3) \left[ (\partial_\mathfrak{c} g)^2 + g \partial^2_\mathfrak{c} g \right].
\end{equation}

It can be noticed that because of the two previous relations, $\mathcal{A}_{4}^{\{ \mathbb{H},\mathbb{D} \}}$ (\ref{anomalie_gauss:H,D 4}) vanishes without any new condition. Then, injecting the previous expression in $\mathcal{A}_{5}^{\{ \mathbb{H},\mathbb{D} \}}=0$ (\ref{anomalie_gauss:H,D 5}) leads to an expression for $\alpha_7$, and, thus, also for $\alpha_8$,

\begin{align}
    \alpha_7=\alpha_8&=4 g \partial_\mathfrak{c}g \left[ (h+\xi_3) + \mathfrak{p}\partial_\mathfrak{p}(h+\xi_3) \right] - \Tilde{g}^2 \nonumber \\& -2 (h+\xi_3)^2 \left[ (\partial_\mathfrak{c}g)^2 + g \partial^2_\mathfrak{c} g \right] \nonumber\\& - \left[ g^2 + 4 \mathfrak{p} g \partial_\mathfrak{p} g \right] \partial_\mathfrak{c} (h+\xi_3).
\end{align}

We next combine the expressions of $\alpha_1$, $\alpha_5$, and $\alpha_6$ in $\mathcal{A}_{3}^{\{ \mathbb{H},\mathbb{H} \}}=0$ (\ref{anomalie_gauss:H,H 3}), and then account that canceling the anomaly $\mathcal{A}_{16}^{\{ \mathbb{H},\mathbb{H} \}}=0$ (\ref{anomalie_gauss:H,H 16}) implies that $\partial_\mathfrak{c}\alpha_3 = 0$. This yields to a relation between $\alpha_3$ and $\alpha_9$,

\begin{equation}
    2 g (\partial_\mathfrak{c}g) \left[ \alpha_3 + 2 \mathfrak{p} \partial_\mathfrak{p} \alpha_3 - \alpha_9  \right]=0.
\end{equation}

Since $g\neq 0$ and $\partial_\mathfrak{c}g \neq 0$, this implies

\begin{equation}\label{eq:equa diff alpha 3 alpha9}
     \alpha_3 + 2 \mathfrak{p} \partial_\mathfrak{p} \alpha_3 = \alpha_9.
\end{equation}

It is worth noticing that, since $\alpha_3$ does not depend on $\mathfrak{c}$, $\alpha_9$ does not either.
Solving this equation leads to

\begin{equation}
    \alpha_3(\mathfrak{p})=\frac{1}{\sqrt{\mathfrak{p}}} \left[ K + \int ^p_1 \frac{\alpha_9(u)}{2 \sqrt{u}} du \right],
\end{equation}

where $K$ is an integration constant.

At this stage, $\beta_{13}$ is still undetermined. To constrain this counterterm, we require the $\{ \mathbb{H}, \mathbb{H} \}$ bracket to be proportional to the full diffeomorphism constraint (see \cite{DeSousa:2024icf}), which implies that the structure coefficients in front of $\mathbb{D}_\mathfrak{g}$ and $\mathbb{D}_\mathfrak{m}$ must be the same, 

\begin{equation}
    (1+\alpha_3)(1+\alpha_5)=(1+\beta_1)(1+\beta_{13}).
\end{equation}

Using the previous results, this leads to 

\begin{equation}
    \beta_{13}=\left[ 1+\alpha_3(\mathfrak{p}) \right] \partial_\mathfrak{c}(g \partial_\mathfrak{c}g) - 1.
\end{equation}

As $\alpha_9(\mathfrak{p})$ does not appear in any other anomaly and is not constrained in any way, it remains a free function of the theory, which determines, together with the explicit expression of the corrections, $\alpha_3$ and, thus, $\beta_{13}$. All the other counterterms of the Hamiltonian constraint are fully determined by the correction functions and the counterterms from the other constraints. To constrain $\xi_3$, $\chi_1$, $\chi_2$, and the corrections, it is important to check the remaining anomalies. \\

$\mathcal{A}_{1}^{\{ \mathbb{D},\mathbb{G} \}}=0$ (\ref{anomalie_gauss:D,G 1}) implies

\begin{equation}\label{eq: l+chi2 en fct de h xi3 et chi1}
    l+\chi_2= \frac{1}{\mathfrak{p}} (h+\xi_3)(\mathfrak{p} + \chi_1).
\end{equation}

As $\beta_{10}$ and $\beta_{12}$ are related to $(h+\xi_3)$, if $(l+\chi_2)$ fulfills Eq. \ref{eq: l+chi2 en fct de h xi3 et chi1}, then $\mathcal{A}_{4}^{\{ \mathbb{H},\mathbb{G} \}}=0$ and $\mathcal{A}_{5}^{\{ \mathbb{H},\mathbb{G} \}}=0$ (\ref{anomalie_gauss:H,G 4} and \ref{anomalie_gauss:H,G 5}) are automatically verified since $\mathcal{A}_{3}^{\{ \mathbb{H},\mathbb{G} \}}=0$ (\ref{anomalie_gauss:H,G 3}) implies $\partial_\mathfrak{c} \chi_1 = 0$.

Using this later condition, Eq. (\ref{eq: l+chi2 en fct de h xi3 et chi1}) and the expressions of $\alpha_4$ and $\alpha_6$ in $\mathcal{A}_1^{\{ \mathbb{H}, \mathbb{G} \}}=0$ (\ref{anomalie_gauss:H,G 1}) lead to

\begin{equation}
    4 g (\partial_\mathfrak{c} g) \left[ \chi_1 - \mathfrak{p} \partial_\mathfrak{p} \chi_1 \right]=0.
\end{equation}

Since, once again, $g\neq 0$ and $\partial_\mathfrak{c} g \neq 0$, this implies

\begin{equation}\label{equa diff chi1}
    \chi_1 - \mathfrak{p} \partial_\mathfrak{p} \chi_1 = 0.
\end{equation}

Recalling that $\chi_1$ is a function of $\mathfrak{p}$ only, this leads to

\begin{equation}\label{eq: expresion chi_1}
    \chi_1 (\mathfrak{p}) = (L-1) \mathfrak{p},
\end{equation}

with $L$ another integration constant. This allows one to rewrite Eq. (\ref{eq: l+chi2 en fct de h xi3 et chi1}) as

\begin{equation}\label{eq:chi_2=... diffeo et gauss}
    \chi_2 = L (h+\xi_3)-l.
\end{equation}

This is an important result of this section. The relevant \textit{effective} corrections that appear at the level of the constraints are the terms $(l+\chi_2)$ and $(h+\xi_3)$, and Eq. (\ref{eq:chi_2=... diffeo et gauss}) shows that they must be proportional one to the other. More details on this proportionality are given in section \ref{sec : cas xi_3=chi_2=0}.\\

Using this in $\mathcal{A}_{2}^{\{ \mathbb{H},\mathbb{G} \}}=0$ (\ref{anomalie_gauss:H,G 2}) directly cancels the anomaly without any new condition on the counterterms or on the correction function. 
For $\mathcal{A}_{18}^{\{ \mathbb{H}, \mathbb{H} \}}$ (\ref{anomalie_gauss:H,H 18}), this implies 

\begin{equation}\label{eq:new relation 1}
    \left[ \Tilde{g} - g \partial_\mathfrak{c} g \right] \left[ 1- \partial_\mathfrak{c} (h+\xi_3) \right]=0.
\end{equation}

To discuss this new condition, two cases must be distinguished:

\begin{itemize}
    \item First, we consider $\Tilde{g} =g \partial_\mathfrak{c} g$. If we take into account the fact that the physical situation corresponds to $\Tilde{g} =g$, this is actually equivalent to having no correction. Equation (\ref{eq:new relation 1}) is then directly satisfied, and $\xi_3(\mathfrak{c},\mathfrak{p})$ is a new degree of freedom, as there is no additional anomaly to constrain it.
    \item The second case is $\Tilde{g} \neq g \partial_\mathfrak{c} g$. Then, Eq. (\ref{eq:new relation 1}) leads to
\begin{equation}\label{eq : xi_3 si g_tilde neq g partial_c g}
    \xi_3(\mathfrak{c},\mathfrak{p})= \mathfrak{c} - h(\mathfrak{c},\mathfrak{p})+ f(\mathfrak{p}),
\end{equation}
with $f(\mathfrak{p})$ an arbitrary function of $\mathfrak{p}$. This function introduces a new degree of freedom into the theory, but it must be chosen so that the correct classical limit of $\xi_3$ is preserved.
\end{itemize}

Finally, $\mathcal{A}_{1}^{\{ \mathbb{H}, \mathbb{H} \}}=0$ (\ref{anomalie_gauss:H,H 1}) leads to a restriction on $g$,

\begin{equation}\label{eq: new relation 2}
    g\partial^2_\mathfrak{c}g+2 \mathfrak{p} \left[ (\partial_\mathfrak{c}g) (\partial_\mathfrak{c}\partial_\mathfrak{p} g)-(\partial_\mathfrak{p} g)(\partial^2_\mathfrak{c} g) \right]=0,
\end{equation}

which is the same condition as the one obtained with $\mathcal{A}_{2}^{\{ \mathbb{H}, \mathbb{H} \}}=0$ (\ref{anomalie_gauss:H,H 2}). Those two anomalies are redundant. This condition is very different from the one obtained in \cite{DeSousa:2024icf,Han:2017wmt}. In particular, Eq. (\ref{eq: new relation 2}) is more general than the condition derived in previous works. 
Solutions presented in those articles were indeed sufficient to cancel the anomalies. However, if one were looking for the most general solution, Eq. (\ref{eq: new relation 2}) -- which is mandatory in our case -- would have been found.\\

It is also worth noting that the standard $\Bar{\mu}$ scheme, where $g=\sin(\Bar{\mu}\mathfrak{c})/\Bar{\mu}$, with $\Bar{\mu}=\sqrt{\Delta l^2_{Pl}/\mathfrak{p}}$, does satisfy Eq. (\ref{eq: new relation 2}).\\

This concludes the closure of the algebra for arbitrary correction functions $g$, $\Tilde{g}$, $h$, and $l$, as the other anomalies are automatically vanishing with the previous conditions. In summary, the theory requires only one condition on the correction functions given by Eq. (\ref{eq: new relation 2}). Specifically, $l$ and $h$ are unconstrained by the algebra as long as the counterterms are included. Furthermore, there are two new degrees of freedom with respect to the case where only the Hamiltonian constraint is corrected \cite{DeSousa:2024icf}.

The first new degree of freedom is the constant $L$ that appears in Eqs. (\ref{eq: expresion chi_1}) and (\ref{eq:chi_2=... diffeo et gauss}). The latter gives the physical interpretation of $L$: it links the \textit{effective} Gauss and diffeomorphism corrections at the level of constraints. In particular, it imposes that $(h+\xi_3)$ must be proportional to $(l+\chi_2)$ -- in other words, that the Gauss and diffeomorphism corrections must be the same, up to a proportionality factor.

The second new degree of freedom is linked to $\xi_3$. As described earlier, in the case where $\Tilde{g}=g\partial_\mathfrak{c} g$, then $\xi_3(\mathfrak{c},\mathfrak{p})$ is a free function of the theory. If this is not the case, then $\xi_3$ is determined by Eq. (\ref{eq : xi_3 si g_tilde neq g partial_c g}), with $f(\mathfrak{p})$ the new degree of freedom, depending only on $\mathfrak{p}$.

One should also notice the $\alpha_9(\mathfrak{p})$ term that is still a free function of the theory together with the constant $K$, as in \cite{DeSousa:2024icf}.\\ 

As a summary, the algebra, at second order in perturbation, for arbitrary correction functions $g$, $\Tilde{g}$, $h$ and $l$ -- if the counterterms fulfill the previous relations (see Appendix \ref{sec:annexe resume contre-termes}) reads

\begin{align}
    \{ \mathbb{G},\mathbb{G}\}=&0,\\
    \{ \mathbb{D},\mathbb{D}\}=&0,\\
    \{ \mathbb{D},\mathbb{G}\}=&0,\\
    \{ \mathbb{H},\mathbb{G}\}=&0,\\
    \{ \mathbb{H}[N],\mathbb{D}[N^a]\}=&\mathbb{H}[\delta N \partial_a \delta N^a],\\
    \{ \mathbb{H}[N_1],\mathbb{H}[N_2]\}=&\frac{1}{2} \partial_\mathfrak{c}^2 g^2 \left[ 1+\alpha_3 \left[ K, \alpha_9 \right] (\mathfrak{p}) \right] \notag\\&\times \mathbb{D} \left[ \frac{{\bf{N}}}{\mathfrak{p}} \partial^a (\delta N_2 - \delta N_1) \right],
\end{align}

with $N={\bf{N}}+\delta N$ the lapse function perturbed around a homogeneous background. 

It is clear that the symplectic structure is mostly unchanged, even with these new degrees of freedom, with respect to \cite{DeSousa:2024icf}.

\subsection{Special cases}

\subsubsection{$\Tilde{g}=g$}

In the case of identical corrections for the background and for the perturbations in the Hamiltonian constraint, Eq. (\ref{eq:new relation 1}) becomes

\begin{equation}
    g [1-\partial_\mathfrak{c} g ] [ 1 - \partial_\mathfrak{c} (h+\xi_3) ]=0.
\end{equation}

To fulfill this condition, either $g=\mathfrak{c}+m_1(\mathfrak{p})$, with $m_1$ an arbitrary function of $\mathfrak{p}$ vanishing at the classical limit, or $\xi_3$ has to fulfill Eq. (\ref{eq : xi_3 si g_tilde neq g partial_c g}). It should be emphasized that the usual correction does not match this requirement, which makes the situation rather ``unnatural". Of course, $g=\Tilde{g}$ must satisfy Eq. (\ref{eq: new relation 2}). The other corrections ($h$ and $l$) remain free. \\

The same conclusions can be obtained in the case $\Tilde{g}=g=h=l$.

\subsubsection{$\Tilde{g}=g=\mathfrak{c}$}

In this case, Eq. (\ref{eq:new relation 1}) is trivially satisfied, as is Eq. (\ref{eq: new relation 2}). There are no restrictions on $h$ and $l$ as long as the counterterms are included. This interestingly shows that correcting the Gauss and diffeomorphism constraint is possible without correcting the Hamiltonian one.

\subsubsection{Case $\xi_3=\chi_2=0$}\label{sec : cas xi_3=chi_2=0}
This subsection focuses on the case where the two counterterms $\xi_3$ and $\chi_2$ are both vanishing. This allows one to emphasize the connection between the different corrections.

In particular, this modifies Eq. (\ref{eq:chi_2=... diffeo et gauss}), leading to 

\begin{equation}\label{eqwithout2counterterms:l prop to h}
    l= L \times h.
\end{equation}

The corrections appearing in the Gauss and diffeomorphism constraints must be proportional one to the other. In particular, in the case where the Gauss constraint is not corrected, Eq. (\ref{eqwithout2counterterms:l prop to h}) leads to $h=\mathfrak{c}/L$. To get the classical limit of $h$, $L$ must be equal to one, leading to $h=\mathfrak{c}$. Thus, there cannot be a diffeomorphism correction without correcting the Gauss constraint (the reciprocal is also true). Furthermore, in this case, $L=1$ implies that $\chi_1=0$ through Eq. \ref{eq: expresion chi_1}.\\

Equation (\ref{eq:new relation 1}) is the only other equation of the previous section for which the physical implications are modified, such that

\begin{equation}
    \left[ \Tilde{g} - g \partial_\mathfrak{c} g \right] \left[ 1- \partial_\mathfrak{c} h \right]=0.
\end{equation}

If $\Tilde{g} \neq g \partial_\mathfrak{c} g$, this implies $h=\mathfrak{c} + m_2(\mathfrak{p})$, where $m_2$ is an arbitrary function of $\mathfrak{p}$, which vanishes at the classical limit ($\mathfrak{p} \rightarrow +\infty$). This leads to Eq. (\ref{eqwithout2counterterms:l prop to h}) implying that $l=\mathfrak{c}+m_2(\mathfrak{p})$, as $L=1$ for the same reason as earlier (leading to $\chi_1=0$).\\

However, if $\Tilde{g} = g \partial_\mathfrak{c} g$, then the diffeomorphism and Gauss constraints are free, as long as they are proportional to each other; see Eq. (\ref{eqwithout2counterterms:l prop to h}). This case is particularly noteworthy since, if $\Tilde{g}=g$, then Eq. (\ref{eq:new relation 1}) yields $g=\mathfrak{c} + m_3(\mathfrak{p})$, with $m_3$ a function of $\mathfrak{p}$ that vanishes at the classical limit.\\


Obviously, in all cases, Eq. (\ref{eq: new relation 2}) must be fulfilled.

\section{Summary and conclusion}

If only the diffeomorphism and Hamiltonian constraints are corrected, the cancellation of anomalies implies the following:

\begin{itemize}
    \item The diffeomorphism correction is absorbed by its associated counterterm, Eq. (\ref{eq_diffeo: h+xi_3 condition}), and leads to no \textit{effective} correction at the level of the constraint\footnote{In previous studies, the philosophy was, in a way, to focus on the correction applied to the constraint and to consider the counterterms as a side effect ensuring consistency. Taking this viewpoint, we could claim here that the diffeomorphism constraint {\it can} be consistently corrected. However, in this specific case, the associated counterterm strictly removes the correction -- this does not happen in usual case where only the Hamiltonian constraint is corrected. This is why we prefer to emphasize in this work that the correction {\it effectively} appearing in the diffeomorphism constraint actually vanishes. Considering that it is ``present" but entirely cancelled by the counterterm would be very artificial.}. Without the counterterm, it is impossible to consistently correct the diffeomorphism constraint.
    
    \item The solution of the system is the same as the one obtained without diffeomorphism correction (see \cite{DeSousa:2024icf}).
    
    \item The condition \begin{equation}\label{eq:relationMaxime g}
        g- 2 \mathfrak{p} \partial_\mathfrak{p} g-\mathfrak{c} \partial_\mathfrak{c} g = 0,
    \end{equation}
    derived in previous works \cite{DeSousa:2024icf,Han:2017wmt}, is sufficient to ensure the cancellation of $\mathcal{A}_{1}^{\{ \mathbb{H}, \mathbb{H} \}}$ and $\mathcal{A}_{2}^{\{ \mathbb{H}, \mathbb{H} \}}$.

\end{itemize}

If all the constraints are corrected, then the cancellation of the anomalies implies the following:

\begin{itemize}
    \item The Gauss and diffeomorphism corrections, added with their associated counterterms (see the previous footnote), must be proportional to each other -- Eq. (\ref{eq:chi_2=... diffeo et gauss}). In particular, without counterterms, the Gauss and diffeomorphism corrections are exactly proportional -- Eq. (\ref{eqwithout2counterterms:l prop to h}). The associated proportionality constant is a new degree of freedom with respect to the previous case which can be fixed in some specific cases.
    
    \item $\xi_3(\mathfrak{c},\mathfrak{p})$ is a new degree of freedom of the theory. If $\Tilde{g}=g\partial_\mathfrak{c} g$, then $\xi_3$ is a fully undetermined function of $\mathfrak{c}$ and $\mathfrak{p}$. However, if $\Tilde{g}\neq g\partial_\mathfrak{c} g$, then $\xi_3$ is partially fixed by the corrections up to a function of $\mathfrak{p}$ (see Eq. (\ref{eq : xi_3 si g_tilde neq g partial_c g})).
    
    \item Equation (\ref{eq:relationMaxime g}) is not enough to get a first-class algebra. It is the condition (\ref{eq: new relation 2}) that needs to be fufilled by $g$.
    
\end{itemize}

In this article, we have shown that it is impossible to correct only the diffeomorphism constraint (assuming, of course, that the Hamiltonian constraint is also corrected, as it should be). The general relativity expression must be maintained.

However, if the Gauss constraint is also modified, a first-class Poisson algebra can be obtained. Interestingly, the differential equations for the correction functions are not the usual ones.\\

This study shows the richness of the deformed algebra approach. In the future, two related investigations should be considered. First, it would be important to understand better the meaning -- at the fundamental level -- of the correction incorporated here at the effective level. Second, it would be interesting to study phenomenological consequences by deriving the gauge-invariant equations for cosmological perturbations and the associated power spectrum.

\bibliography{refs.bib}
\onecolumngrid
\appendix
\newpage

\section{Expression of constraints when all corrected}\label{sec:appendix_expconstraints}
\numberwithin{equation}{section}

First, the Hamiltonian constraint reads

\begin{align}
    \mathbb{H} \left[N\right] &= \mathbb{H}_\mathfrak{g}\left[N\right] + \mathbb{H}_\mathfrak{m}\left[N\right] \notag\\
    &=\frac{1}{2 \kappa} \int d{\bf{x}} \; \big( {\bf{N}} \big[ \mathcal{H}_\mathfrak{g}^{(0)} + \mathcal{H}_\mathfrak{g}^{(2)} \big] + \delta N \mathcal{H}_\mathfrak{g}^{(1)} \big) \notag\\
    &+\frac{1}{2 \kappa} \int d{\bf{x}} \; \big( {\bf{N}} \big[ \mathcal{H}_\mathfrak{m}^{(0)} + \mathcal{H}_\mathfrak{m}^{(2)} \big] + \delta N \mathcal{H}_\mathfrak{m}^{(1)} \big),
\end{align}

with, for the geometrical sector, at the background level,

\begin{equation}
    \mathcal{H}_\mathfrak{g}^{(0)} = - 6 \sqrt{\mathfrak{p}} g^2.
\end{equation}

The first-order term of the geometrical sector reads

\begin{equation}
    \mathcal{H}_\mathfrak{g}^{(1)} = - 4 \sqrt{\mathfrak{p}} (\Tilde{g}+\alpha_1) \delta K _b^b - \frac{1}{\sqrt{\mathfrak{p}}} (\Tilde{g}^2+\alpha_2) \delta E_b^b + \frac{2}{\sqrt{\mathfrak{p}}} (1+\alpha_3) \partial_a \partial^i \delta E_i^a,
\end{equation}

while the second-order one is

\begin{align}
    \mathcal{H}_\mathfrak{g}^{(2)}&=\sqrt{\mathfrak{p}}  (1+\alpha_4) \delta K_a^b \delta K^b_a - \sqrt{\mathfrak{p}}(1+\alpha_5)(\delta K_b^b)^2  \notag\\
    &- \frac{2}{\sqrt{\mathfrak{p}}} (\Tilde{g}+\alpha_6)  \delta K^i_a \delta E^a_i  - \frac{1}{2 \mathfrak{p}^{3/2}} (\Tilde{g}^2+\alpha_7) \delta E_b^a \delta E _a^b \notag\\
    &+ \frac{1}{4 \mathfrak{p}^{3/2}}   (\Tilde{g}^2+\alpha_8) (\delta E_b^b)^2 + \frac{1}{\mathfrak{p}^{3/2}} (1+\alpha_9) \mathcal{Z}^{cidj}_{ab}  ( \partial_c \delta E^a_i)(\partial_d \delta E^b_j),
\end{align}

where

\begin{equation}
    \mathcal{Z}^{cidj}_{ab}=\frac{1}{4} \epsilon_{k}{}^{ef}\epsilon_{mn}{}^{k} \mathcal{X}^{mjd}_{be} \mathcal{X}^{nic}_{af}-\epsilon_{k}{}^{ie}\mathcal{X}^{kjd}_{be}\delta^c_a-\epsilon_{k}{}^{ci} \mathcal{X}^{kjd}_{ba} + \frac{1}{2} \delta_a^i \epsilon_{k}{}^{ce} \mathcal{X}^{kjd}_{be},
\end{equation}

with $\epsilon^{abc}$ the Levi-Civita symbol, $\delta^a_b$ the Kronnecker delta, while

\begin{equation}
    \mathcal{X}^{ijb}_{ca}=\epsilon_{c}{}^{ij} \delta_a^b - \epsilon_{c}{}^{ib} \delta^j_a + \epsilon^{ijb}\delta_{ca}+\epsilon_{a}{}^{ib} \delta^j_c.
\end{equation}

For the matter sector of $\mathbb{H}$, the background contribution reads

\begin{equation}
    \mathcal{H}_\mathfrak{m}^{(0)}=\frac{\bm{\pi}^2}{2 \mathfrak{p}^{3/2}} + \mathfrak{p}^{3/2} V[\bm{\phi}],
\end{equation}

with $\bm{\phi}$ and $\bm{\pi}$ the matter field (with potential $V[\bm{\phi}]$) and its conjugate momentum at the background level. Introducing the perturbation of the scalar field $\delta \phi$ and its conjugate momentum $\delta \pi$, such that $\phi = \bm{\phi}+\delta \phi$ and $\pi = \bm{\pi}+ \delta \pi$, the first-order perturbations of the matter sector read

\begin{equation}
    \mathcal{H}_\mathfrak{m}^{(1)} = \frac{\bm{\pi}}{\mathfrak{p}^{3/2}} (1+\beta_1) \delta \pi - \frac{\bm{\pi}^2}{4 \mathfrak{p}^{5/2}} (1+\beta_2)  \delta E^a_a + \mathfrak{p}^{3/2} (\partial_\phi V[\bm{\phi}])  (1+\beta_3) \delta \phi +  \frac{\sqrt{\mathfrak{p}}}{2} V[\bm{\phi}] (1+\beta_4)  \delta E^a_a.
\end{equation}

The second-order term is

\begin{align}
    \mathcal{H}_\mathfrak{m}^{(2)}&=\frac{1}{2\mathfrak{p}^{3/2}} (1+\beta_5) (\delta \pi)^2 - \frac{\bm{\pi}}{2 \mathfrak{p}^{5/2}} (1+\beta_6) \delta \pi \delta E_a^a \notag \\
    &+ \frac{\mathfrak{p}^{3/2}}{2}  (\partial_\phi^2 V[\bm{\phi}]) (1+\beta_7) (\delta \phi)^2 + \frac{\sqrt{\mathfrak{p}}}{2} (\partial_\phi V[\bm{\phi}]) (1+\beta_8) \delta \phi \delta E^a_a\notag\\
    &+ \frac{\bm{\pi}^2}{16 \mathfrak{p}^{7/2}} (1+\beta_9)  (\delta E^a_a)^2 + \frac{\bm{\pi}^2}{8 \mathfrak{p}^{7/2}} (1+\beta_{10})  \delta E^a_b \delta E^b_a \notag\\
    &+ \frac{V[\bm{\phi}]}{8 \sqrt{\mathfrak{p}}} (1+\beta_{11})  (\delta E^a_a)^2 - \frac{V[\bm{\phi}]}{4 \sqrt{\mathfrak{p}}} (1+\beta_{12}) \delta E^a_b \delta E^b_a\notag\\
    &+ \frac{\sqrt{\mathfrak{p}}}{2} (1+\beta_{13}) (\partial_a \delta \phi)(\partial^a \delta \phi).
\end{align}

For the diffeomorphism constraint, one gets

\begin{align}
    \mathbb{D} \left[N^a\right] &= \mathbb{D}_\mathfrak{g} \left[N^a\right] + \mathbb{D}_\mathfrak{m} \left[N^a\right]\notag\\
    &=\frac{1}{\kappa} \int d{\bf{x}} \; \delta N^a \mathcal{D}^{\mathfrak{g}(1)}_a + \frac{1}{\kappa} \int d{\bf{x}} \; \delta N^a \mathcal{D}^{\mathfrak{m}(1)}_a  ,
\end{align}

where

\begin{equation}
    \mathcal{D}^{\mathfrak{g}(1)}_a =\mathfrak{p}(1+\xi_1)\partial_a \delta K_b^b-\mathfrak{p}(1+\xi_2)\partial_i \delta K^i_a - (h+\xi_3) \delta^j_a \partial_b \delta E^b_j,
\end{equation}

with $h=h(\mathfrak{c},\mathfrak{p})$ the correction of the curvature $\mathfrak{c}$ in the diffeomorphism constraint. The matter sector reads

\begin{equation}
    \mathcal{D}^{\mathfrak{m}(1)}_a ={\bm{\pi}} \partial_a \delta \phi.
\end{equation}

Finally, the Gauss constraint is given by

\begin{equation}
    \mathbb{G} \left[\Lambda^i\right]=\frac{1}{\kappa} \int d{\bf{x}} \; \delta \Lambda^i \mathcal{G}_i^{(1)},
\end{equation}

with

\begin{equation}
     \mathcal{G}_i = (\mathfrak{p}+\chi_1) \varepsilon_{ij}^a \delta K_a^j + (l+\chi_2) \varepsilon_{ia}^j \delta E^a_j.
\end{equation}

In section \ref{sec:correc diffeo}, the counterterms $\chi_1$, $\chi_2$, and $\chi_3$ are set to zero, while $l=\mathfrak{c}$.

\section{List of anomalies without correcting the Gauss constraint}\label{sec:list anomalies juste diffeo}

\numberwithin{equation}{section}

There are no anomalies for the $\left\{ \mathbb{G},\mathbb{G} \right\}$ and $\left\{ \mathbb{D},\mathbb{D} \right\}$ brackets. For the $\left\{ \mathbb{D},\mathbb{G} \right\}$ bracket one anomaly appears :

\begin{equation}
    \mathcal{A}_1^{\left\{ \mathbb{D},\mathbb{G} \right\}} = (1+\xi_2) \mathfrak{c} -h-\xi_3.\label{anomalie_diffeo:D,G 1}
\end{equation}
\newpage
For the $\left\{ \mathbb{H},\mathbb{G} \right\}$ bracket,

\begin{align}
    &\mathcal{A}_1^{\left\{ \mathbb{H},\mathbb{G} \right\}} = g^2 - 2 \mathfrak{c}(\Tilde{g}+\alpha_6)+(\Tilde{g}^2+\alpha_7) +4 \mathfrak{p} g \partial_\mathfrak{p} g \label{anomalie_diffeo:H,G 1}\\
    &\mathcal{A}_2^{\left\{ \mathbb{H},\mathbb{G} \right\}} =  \mathfrak{c} (1+\alpha_4) +(\Tilde{g}+\alpha_6)-2 g \partial_\mathfrak{c} g \label{anomalie_diffeo:H,G 2}\\
    &\mathcal{A}_3^{\left\{ \mathbb{H},\mathbb{G} \right\}} = - \beta_{10} \label{anomalie_diffeo:H,G 3}\\
    &\mathcal{A}_4^{\left\{ \mathbb{H},\mathbb{G} \right\}} =  \beta_{12} \label{anomalie_diffeo:H,G 4}.
\end{align}

Then for the $\left\{ \mathbb{H},\mathbb{D} \right\}$ bracket,

\begin{align}
    &\mathcal{A}_1^{\left\{ \mathbb{H},\mathbb{D} \right\}}=(\Tilde{g}^2+\alpha_2)(\xi_2 - 2 -3 \xi_1)-4(\Tilde{g}+\alpha_1)(h+\xi_3) + 6 g^2 \label{anomalie_diffeo:H,D 1}\\
    &\mathcal{A}_2^{\left\{ \mathbb{H},\mathbb{D} \right\}}=(1+\alpha_3)(\xi_1-\xi_2)\label{anomalie_diffeo:H,D 2}\\
    &\mathcal{A}_3^{\left\{ \mathbb{H},\mathbb{D} \right\}}=2(1+\alpha_4)(h+\xi_3)+2(\Tilde{g}+\alpha_6)(1+\xi_2) - 4g (1+\xi_2 + \mathfrak{p} \partial_\mathfrak{p} \xi_2 ) \partial_\mathfrak{c}g + (g^2 + 4 \mathfrak{p} g \partial_\mathfrak{p} g ) \partial_\mathfrak{c} \xi_2 \label{anomalie_diffeo:H,D 3}\\
    &\mathcal{A}_4^{\left\{ \mathbb{H},\mathbb{D} \right\}}= 4 g(1+\xi_1+\mathfrak{p} \partial_\mathfrak{p} \xi_1 ) \partial_\mathfrak{c} g - (g^2 + 4\mathfrak{p}g \partial_\mathfrak{p} g) \partial_\mathfrak{c} \xi_1 - 2 (1+\alpha_5)(h+\xi_3) - 2(\Tilde{g} + \alpha_6)(1+\xi_1) \label{anomalie_diffeo:H,D 4}\\
    &\mathcal{A}_5^{\left\{ \mathbb{H},\mathbb{D} \right\}}=(\Tilde{g}^2 + \alpha_7)(1+\xi_2)-2(\Tilde{g} + \alpha_6)(h+\xi_3) - 4 \mathfrak{p} g (\partial_\mathfrak{c} g) \partial_\mathfrak{p}(h+\xi_3) + (g^2+4\mathfrak{p} g \partial_\mathfrak{p}g)\partial_\mathfrak{c}(h+\xi_3) \label{anomalie_diffeo:H,D 5}\\
    &\mathcal{A}_6^{\left\{ \mathbb{H},\mathbb{D} \right\}}=  (\Tilde{g}^2+\alpha_8)(2+3\xi_1-\xi_2)-2(\Tilde{g}^2+\alpha_7)(1+\xi_1) \label{anomalie_diffeo:H,D 6}\\
    &\mathcal{A}_7^{\left\{ \mathbb{H},\mathbb{D} \right\}}=(1+\alpha_9)(\xi_1-\xi_2). \label{anomalie_diffeo:H,D 7}\\
    &\mathcal{A}_8^{\left\{ \mathbb{H},\mathbb{D} \right\}}= \partial_\mathfrak{c} \xi_1 \label{anomalie_diffeo:H,D 8}\\
    &\mathcal{A}_9^{\left\{ \mathbb{H},\mathbb{D} \right\}}= \partial_\mathfrak{c} \xi_2 \label{anomalie_diffeo:H,D 9}\\
    &\mathcal{A}_{10}^{\left\{ \mathbb{H},\mathbb{D} \right\}}= 1+ 2 \beta_1 -(1+\xi_1)(1+\beta_2) \label{anomalie_diffeo:H,D 10}\\
    &\mathcal{A}_{11}^{\left\{ \mathbb{H},\mathbb{D} \right\}}= 1+\beta_5 - (1+\xi_1)(1+\beta_6) \label{anomalie_diffeo:H,D 11}\\
    &\mathcal{A}_{12}^{\left\{ \mathbb{H},\mathbb{D} \right\}}=  \partial_\mathfrak{c} (h+\xi_3) - (1+\xi_2)(1+\beta_{10})\label{anomalie_diffeo:H,D 12}\\
    &\mathcal{A}_{13}^{\left\{ \mathbb{H},\mathbb{D} \right\}}= (1+\xi_2)(1+\beta_{12}) - \partial_\mathfrak{c}(h+\xi_3) \label{anomalie_diffeo:H,D 13}\\
    &\mathcal{A}_{14}^{\left\{ \mathbb{H},\mathbb{D} \right\}}= (1+\xi_1)(2+\beta_{10}+ \beta_9) - 2 - 2 \beta_6 \label{anomalie_diffeo:H,D 14}\\
    &\mathcal{A}_{15}^{\left\{ \mathbb{H},\mathbb{D} \right\}}= (1+\xi_1)(\beta_{11}-\beta_{12}) \label{anomalie_diffeo:H,D 15}\\
    &\mathcal{A}_{16}^{\left\{ \mathbb{H},\mathbb{D} \right\}}= (1+\xi_1)(1+\beta_8)-1\label{anomalie_diffeo:H,D 16}\\
    &\mathcal{A}_{17}^{\left\{ \mathbb{H},\mathbb{D} \right\}}= (1+\xi_1)(1+\beta_4)-1.\label{anomalie_diffeo:H,D 17}
\end{align}

Finally, the $\left\{ \mathbb{H},\mathbb{H} \right\}$ bracket,

\begin{align}
    \mathcal{A}_1^{\left\{ \mathbb{H},\mathbb{H} \right\}}=& 4(\Tilde{g} + \alpha_1) \left( g \partial_\mathfrak{c} g - \alpha_6 - \Tilde{g} \right) - 2 ( g^2 + 4 g \mathfrak{p} \partial_\mathfrak{p} g) \partial_\mathfrak{c} ( \Tilde{g}+\alpha_1) + 8 \mathfrak{p} g (\partial_\mathfrak{c} g) \partial_\mathfrak{p}(\Tilde{g}+\alpha_1) \notag \\
    &+ (\Tilde{g}^2 + \alpha_2)(2+3\alpha_5 - \alpha_4)\label{anomalie_diffeo:H,H 1}\\
    \mathcal{A}_2^{\left\{ \mathbb{H},\mathbb{H} \right\}}=& 2 (\Tilde{g}^2 + \alpha_2)(\alpha_6 - g \partial_\mathfrak{c}g + \Tilde{g}) -(g^2 + 4 \mathfrak{p} g \partial_\mathfrak{p} g) \partial_\mathfrak{c} (\Tilde{g}^2 + \alpha_2) + 4  \mathfrak{p} g (\partial_\mathfrak{c} g) \partial_\mathfrak{p} (\Tilde{g}^2 + \alpha_2)\notag \\
    &+ 2 ( \Tilde{g} + \alpha_1) ( \Tilde{g}^2 - 2 \alpha_7 + 3 \alpha_8) \label{anomalie_diffeo:H,H 2}\\
    \mathcal{A}_3^{\left\{ \mathbb{H},\mathbb{H} \right\}}=& 2(1+\alpha_3) \left[ \Tilde{g} + \alpha_6 - g \partial_\mathfrak{c} g + (h+\xi_3)(1+\alpha_5)\right] -2 (\Tilde{g}+\alpha_1)(1+\alpha_9) + 4 \mathfrak{p} g (\partial_\mathfrak{p} \alpha_3)(\partial_\mathfrak{c} g) \notag \\
    &- (g^2 + 4 \mathfrak{p} g \partial_\mathfrak{p} g) \partial_\mathfrak{c} \alpha_3 \label{anomalie_diffeo:H,H 3}\\
    \mathcal{A}_4^{\left\{ \mathbb{H},\mathbb{H} \right\}}=& (1+\alpha_3) \left[ (1+\xi_2)(1+\alpha_5) -1-\alpha_4 \right] \label{anomalie_diffeo:H,H 4}\\
    \mathcal{A}_5^{\left\{ \mathbb{H},\mathbb{H} \right\}}=& - \xi_1 (1+\alpha_3)(1+\alpha_5)\label{anomalie_diffeo:H,H 5}\\
    \mathcal{A}_6^{\left\{ \mathbb{H},\mathbb{H} \right\}}=& \partial_\mathfrak{c} \beta_1 \label{anomalie_diffeo:H,H 6}
\end{align}
\begin{align}
    \mathcal{A}_7^{\left\{ \mathbb{H},\mathbb{H} \right\}}=& \partial_\mathfrak{c} \beta_3 \label{anomalie_diffeo:H,H 7}\\
    \mathcal{A}_8^{\left\{ \mathbb{H},\mathbb{H} \right\}}=& -\partial_\mathfrak{c} \beta_2 \label{anomalie_diffeo:H,H 8}\\
    \mathcal{A}_9^{\left\{ \mathbb{H},\mathbb{H} \right\}}=& -\partial_\mathfrak{c} \beta_4 \label{anomalie_diffeo:H,H 9}\\
    \mathcal{A}_{10}^{\left\{ \mathbb{H},\mathbb{H} \right\}}=& \partial_\mathfrak{c}( \beta_2 + \beta_4) \label{anomalie_diffeo:H,H 10}\\
    \mathcal{A}_{11}^{\left\{ \mathbb{H},\mathbb{H} \right\}}=& \beta_1 - \beta_3 - \beta_5 - \beta_3 \beta_5\label{anomalie_diffeo:H,H 11}\\
    \mathcal{A}_{12}^{\left\{ \mathbb{H},\mathbb{H} \right\}}=& \beta_1 + \beta_7 + \beta_1 \beta_7 - \beta_3\label{anomalie_diffeo:H,H 12}\\
    \mathcal{A}_{13}^{\left\{ \mathbb{H},\mathbb{H} \right\}}=&  \beta_1 + \beta_3 + \beta_6 + \beta_3 \beta_6 + \beta_8 + \beta_1 \beta_8 - \beta_2 - \beta_4 \label{anomalie_diffeo:H,H 13}\\
    \mathcal{A}_{14}^{\left\{ \mathbb{H},\mathbb{H} \right\}}= &\left[ 1 + \beta_2 \right] (2 - \alpha_4 + 3 \alpha_5) - 2 \partial_\mathfrak{c} (\Tilde{g} + \alpha_1) \label{anomalie_diffeo:H,H 14}\\
    \mathcal{A}_{15}^{\left\{ \mathbb{H},\mathbb{H} \right\}}=& \left[ 1 + \beta_4 \right] (-2 + \alpha_4 - 3 \alpha_5) + 2 \partial_\mathfrak{c} (\Tilde{g} + \alpha_1)\label{anomalie_diffeo:H,H 15} \\
    \mathcal{A}_{16}^{\left\{ \mathbb{H},\mathbb{H} \right\}}=& \partial_\mathfrak{c} \alpha_3 \label{anomalie_diffeo:H,H 16}\\
    \mathcal{A}_{17}^{\left\{ \mathbb{H},\mathbb{H} \right\}}=& -6 \left[ \Tilde{g} + \alpha_1 \right] \left[ 1 +\beta_6 \right] + 2 g (\partial_\mathfrak{c}g)(3 +3 \beta_1 - 2 \mathfrak{p} \partial_\mathfrak{p} \beta_1) + g^2 \partial_\mathfrak{c} \beta_1 + 4 g \mathfrak{p} (\partial_\mathfrak{p} g)(\partial_\mathfrak{c} \beta_1) \label{anomalie_diffeo:H,H 17}\\
    \mathcal{A}_{18}^{\left\{ \mathbb{H},\mathbb{H} \right\}}=& 12 \Tilde{g} + 10 \alpha_1 + 2 \alpha_6 + 4(\Tilde{g} + \alpha_1) \beta_{10} + 2 (\Tilde{g} + \alpha_6) \beta_2 + 6 \Tilde{g} \beta_9 - 2 \Tilde{g} \partial_\mathfrak{c} \Tilde{g} - 2 g (\partial_\mathfrak{c} g)(5+5 \beta_2 - 2 \mathfrak{p} \partial_\mathfrak{p} \beta_2) \notag\\
    &- \partial_\mathfrak{c} \alpha_2 - g^2 \partial_\mathfrak{c} \beta_2 - 4 g \mathfrak{p} (\partial_\mathfrak{p} g)(\partial_\mathfrak{c} \beta_2) \label{anomalie_diffeo:H,H 18}\\
    \mathcal{A}_{19}^{\left\{ \mathbb{H},\mathbb{H} \right\}}= &2 \alpha_1 - 2 \alpha_6 + 6 (\Tilde{g} + \alpha_1) \beta_{11} - 4 ( \Tilde{g} + \alpha_1) \beta_{12} - 2 (\Tilde{g} + \alpha_6) \beta_4 + 2 \Tilde{g} \partial_\mathfrak{c} \Tilde{g} -2 g (\partial_\mathfrak{c} g)(1+\beta_4 + 2 \mathfrak{p} \partial_\mathfrak{p} \beta_4) \notag\\
    &+ \partial_\mathfrak{c} \alpha_2 +g^2 \partial_\mathfrak{c} \beta_4 + 4 g \mathfrak{p} (\partial_\mathfrak{p} g)(\partial_\mathfrak{c} \beta_4) \label{anomalie_diffeo:H,H 19}\\
    \mathcal{A}_{20}^{\left\{ \mathbb{H},\mathbb{H} \right\}}= &3 \mathfrak{p} (\Tilde{g} + \alpha_1) + 3 \mathfrak{p} (\Tilde{g} + \alpha_1)\beta_8 - g (\partial_\mathfrak{c}g)\left[ 3 \mathfrak{p} (1+\beta_3) + 2 \mathfrak{p}^2 \partial_\mathfrak{p} \beta_3 \right] + \frac{\mathfrak{p}}{2} g^2 \partial_\mathfrak{c} \beta_3 + 2 \mathfrak{p}^2 g (\partial_\mathfrak{p} g)(\partial_\mathfrak{c} \beta_3).\label{anomalie_diffeo:H,H 20} 
\end{align}

\section{List of anomalies when correcting every constraints}\label{sec:list anomalies tout corrigé}
\numberwithin{equation}{section}

In this case, there are no anomalies for the $\left\{ \mathbb{G},\mathbb{G} \right\}$ and $\left\{ \mathbb{D},\mathbb{D} \right\}$ brackets. For the $\left\{ \mathbb{D},\mathbb{G} \right\}$ bracket, there is one anomaly :

\begin{align}
    \mathcal{A}_1^{\left\{ \mathbb{D},\mathbb{G} \right\}} =& \mathfrak{p}(1+\xi_2)(l+\chi_2) -(h+\xi_3)(\mathfrak{p}+\chi_1).\label{anomalie_gauss:D,G 1}
\end{align}

For the $\left\{ \mathbb{H},\mathbb{G} \right\}$ bracket, the anomalies read

\begin{align}
    \mathcal{A}^{\{ \mathbb{H},\mathbb{G}\}}_1=&g^2 \partial_\mathfrak{c} \chi_1 + 4 \mathfrak{p} g \left[ (\partial_\mathfrak{p} g)(\partial_\mathfrak{c} \chi_1) - (\partial_\mathfrak{c} g)(1+\partial_\mathfrak{p} \chi_1) \right] + 2 \mathfrak{p}(1+\alpha_4)(l+\chi_2) + 2 (\Tilde{g}+\alpha_6)(\mathfrak{p}+\chi_1)\label{anomalie_gauss:H,G 1} \\
    \mathcal{A}^{\{ \mathbb{H},\mathbb{G}\}}_2=&(\Tilde{g}^2+\alpha_7)(\mathfrak{p}+\chi_1) - 2 \mathfrak{p} (\Tilde{g}+\alpha_6)(l+\chi_2) + \mathfrak{p} g^2 \partial_\mathfrak{c}(l+\chi_2) + 4 \mathfrak{p}^2 g \left[ (\partial_\mathfrak{p} g)\partial_\mathfrak{c}(l+\chi_2) - (\partial_\mathfrak{c} g) \partial_\mathfrak{p}(l+\chi_2) \right]\label{anomalie_gauss:H,G 2}\\
    \mathcal{A}^{\{ \mathbb{H},\mathbb{G}\}}_3=&\partial_\mathfrak{c} \chi_1\label{anomalie_gauss:H,G 3}\\
    \mathcal{A}^{\{ \mathbb{H},\mathbb{G}\}}_4=&\mathfrak{p}\partial_\mathfrak{c}(l+\chi_2)-(\mathfrak{p}+\chi_1)(1+\beta_{10})\label{anomalie_gauss:H,G 4}\\
    \mathcal{A}^{\{ \mathbb{H},\mathbb{G}\}}_5=&(\mathfrak{p} + \chi_1)(1+\beta_{12})-\mathfrak{p} \partial_\mathfrak{c}(l+\chi_2).\label{anomalie_gauss:H,G 5} 
\end{align}

For the $\left\{ \mathbb{H},\mathbb{D} \right\}$ bracket,

\begin{align}
&\mathcal{A}_1^{\left\{ \mathbb{H},\mathbb{D} \right\}}=(\Tilde{g}^2+\alpha_2)(\xi_2 - 2 -3 \xi_1)-4(\Tilde{g}+\alpha_1)(h+\xi_3) + 6 g^2 \label{anomalie_gauss:H,D 1}\\
    &\mathcal{A}_2^{\left\{ \mathbb{H},\mathbb{D} \right\}}=(1+\alpha_3)(\xi_1-\xi_2)\label{anomalie_gauss:H,D 2}\\
    &\mathcal{A}_3^{\left\{ \mathbb{H},\mathbb{D} \right\}}=2(1+\alpha_4)(h+\xi_3)+2(\Tilde{g}+\alpha_6)(1+\xi_2) - 4g (1+\xi_2 + \mathfrak{p} \partial_\mathfrak{p} \xi_2 ) \partial_\mathfrak{c}g + (g^2 + 4 \mathfrak{p} g \partial_\mathfrak{p} g ) \partial_\mathfrak{c} \xi_2 \label{anomalie_gauss:H,D 3} \\
    &\mathcal{A}_4^{\left\{ \mathbb{H},\mathbb{D} \right\}}= 4 g(1+\xi_1+\mathfrak{p} \partial_\mathfrak{p} \xi_1 ) \partial_\mathfrak{c} g - (g^2 + 4\mathfrak{p}g \partial_\mathfrak{p} g) \partial_\mathfrak{c} \xi_1 - 2 (1+\alpha_5)(h+\xi_3) - 2(\Tilde{g} + \alpha_6)(1+\xi_1)\label{anomalie_gauss:H,D 4} \\
    &\mathcal{A}_5^{\left\{ \mathbb{H},\mathbb{D} \right\}}=(\Tilde{g}^2 + \alpha_7)(1+\xi_2)-2(\Tilde{g} + \alpha_6)(h+\xi_3) - 4 \mathfrak{p} g (\partial_\mathfrak{c} g) \partial_\mathfrak{p}(h+\xi_3) + (g^2+4\mathfrak{p} g \partial_\mathfrak{p}g)\partial_\mathfrak{c}(h+\xi_3)\label{anomalie_gauss:H,D 5} 
\end{align}
\begin{align}
    &\mathcal{A}_6^{\left\{ \mathbb{H},\mathbb{D} \right\}}=  (\Tilde{g}^2+\alpha_8)(2+3\xi_1-\xi_2)-2(\Tilde{g}^2+\alpha_7)(1+\xi_1) \label{anomalie_gauss:H,D 6}\\
    &\mathcal{A}_7^{\left\{ \mathbb{H},\mathbb{D} \right\}}=(1+\alpha_9)(\xi_1-\xi_2).\label{anomalie_gauss:H,D 7} \\
    &\mathcal{A}_8^{\left\{ \mathbb{H},\mathbb{D} \right\}}= \partial_\mathfrak{c} \xi_1 \label{anomalie_gauss:H,D 8}\\
    &\mathcal{A}_9^{\left\{ \mathbb{H},\mathbb{D} \right\}}= \partial_\mathfrak{c} \xi_2 \label{anomalie_gauss:H,D 9}\\
    &\mathcal{A}_{10}^{\left\{ \mathbb{H},\mathbb{D} \right\}}= 1+ 2 \beta_1 -(1+\xi_1)(1+\beta_2) \label{anomalie_gauss:H,D 10}\\
    &\mathcal{A}_{11}^{\left\{ \mathbb{H},\mathbb{D} \right\}}= 1+\beta_5 - (1+\xi_1)(1+\beta_6) \label{anomalie_gauss:H,D 11}\\
    &\mathcal{A}_{12}^{\left\{ \mathbb{H},\mathbb{D} \right\}}=  \partial_\mathfrak{c} (h+\xi_3) - (1+\xi_2)(1+\beta_{10})\label{anomalie_gauss:H,D 12}\\
    &\mathcal{A}_{13}^{\left\{ \mathbb{H},\mathbb{D} \right\}}= (1+\xi_2)(1+\beta_{12}) - \partial_\mathfrak{c}(h+\xi_3) \label{anomalie_gauss:H,D 13}\\
    &\mathcal{A}_{14}^{\left\{ \mathbb{H},\mathbb{D} \right\}}= (1+\xi_1)(2+\beta_{10}+ \beta_9) - 2 - 2 \beta_6 \label{anomalie_gauss:H,D 14}\\
    &\mathcal{A}_{15}^{\left\{ \mathbb{H},\mathbb{D} \right\}}= (1+\xi_1)(\beta_{11}-\beta_{12}) \label{anomalie_gauss:H,D 15}\\
    &\mathcal{A}_{16}^{\left\{ \mathbb{H},\mathbb{D} \right\}}= (1+\xi_1)(1+\beta_8)-1\label{anomalie_gauss:H,D 16}\\
    &\mathcal{A}_{17}^{\left\{ \mathbb{H},\mathbb{D} \right\}}= (1+\xi_1)(1+\beta_4)-1.\label{anomalie_gauss:H,D 17}
\end{align}

Finally for the $\left\{ \mathbb{H},\mathbb{H} \right\}$ bracket,

\begin{align}
    \mathcal{A}_1^{\left\{ \mathbb{H},\mathbb{H} \right\}}=& 4(\Tilde{g} + \alpha_1) \left( g \partial_\mathfrak{c} g - \alpha_6 - \Tilde{g} \right) - 2 ( g^2 + 4 g \mathfrak{p} \partial_\mathfrak{p} g) \partial_\mathfrak{c} ( \Tilde{g}+\alpha_1) + 8 \mathfrak{p} g (\partial_\mathfrak{c} g) \partial_\mathfrak{p}(\Tilde{g}+\alpha_1) \notag \\ &+ (\Tilde{g}^2 + \alpha_2)(2+3\alpha_5 - \alpha_4)\label{anomalie_gauss:H,H 1}\\
    \mathcal{A}_2^{\left\{ \mathbb{H},\mathbb{H} \right\}}=& 2 (\Tilde{g}^2 + \alpha_2)(\alpha_6 - g \partial_\mathfrak{c}g + \Tilde{g}) -(g^2 + 4 \mathfrak{p} g \partial_\mathfrak{p} g) \partial_\mathfrak{c} (\Tilde{g}^2 + \alpha_2) + 4  \mathfrak{p} g (\partial_\mathfrak{c} g) \partial_\mathfrak{p} (\Tilde{g}^2 + \alpha_2)\notag \\&+ 2 ( \Tilde{g} + \alpha_1) ( \Tilde{g}^2 - 2 \alpha_7 + 3 \alpha_8) \label{anomalie_gauss:H,H 2}\\
    \mathcal{A}_3^{\left\{ \mathbb{H},\mathbb{H} \right\}}=& 2(1+\alpha_3) \left[ \Tilde{g} + \alpha_6 - g \partial_\mathfrak{c} g + (h+\xi_3)(1+\alpha_5)\right] -2 (\Tilde{g}+\alpha_1)(1+\alpha_9) + 4 \mathfrak{p} g (\partial_\mathfrak{p} \alpha_3)(\partial_\mathfrak{c} g) \notag \\&- (g^2 + 4 \mathfrak{p} g \partial_\mathfrak{p} g) \partial_\mathfrak{c} \alpha_3 \label{anomalie_gauss:H,H 3}\\
    \mathcal{A}_4^{\left\{ \mathbb{H},\mathbb{H} \right\}}=& (1+\alpha_3) \left[ (1+\xi_2)(1+\alpha_5) -1-\alpha_4 \right] \label{anomalie_gauss:H,H 4}\\
    \mathcal{A}_5^{\left\{ \mathbb{H},\mathbb{H} \right\}}=& - \xi_1 (1+\alpha_3)(1+\alpha_5)\label{anomalie_gauss:H,H 5}\\
    \mathcal{A}_6^{\left\{ \mathbb{H},\mathbb{H} \right\}}=& \partial_\mathfrak{c} \beta_1 \label{anomalie_gauss:H,H 6}\\
    \mathcal{A}_7^{\left\{ \mathbb{H},\mathbb{H} \right\}}=& \partial_\mathfrak{c} \beta_3 \label{anomalie_gauss:H,H 7}\\
    \mathcal{A}_8^{\left\{ \mathbb{H},\mathbb{H} \right\}}=& -\partial_\mathfrak{c} \beta_2 \label{anomalie_gauss:H,H 8}\\
    \mathcal{A}_9^{\left\{ \mathbb{H},\mathbb{H} \right\}}=& -\partial_\mathfrak{c} \beta_4 \label{anomalie_gauss:H,H 9}\\
    \mathcal{A}_{10}^{\left\{ \mathbb{H},\mathbb{H} \right\}}=& \partial_\mathfrak{c}( \beta_2 + \beta_4)\label{anomalie_gauss:H,H 10} \\
    \mathcal{A}_{11}^{\left\{ \mathbb{H},\mathbb{H} \right\}}=& \beta_1 - \beta_3 - \beta_5 - \beta_3 \beta_5\label{anomalie_gauss:H,H 11}\\
    \mathcal{A}_{12}^{\left\{ \mathbb{H},\mathbb{H} \right\}}=& \beta_1 + \beta_7 + \beta_1 \beta_7 - \beta_3\label{anomalie_gauss:H,H 12}\\
    \mathcal{A}_{13}^{\left\{ \mathbb{H},\mathbb{H} \right\}}=&  \beta_1 + \beta_3 + \beta_6 + \beta_3 \beta_6 + \beta_8 + \beta_1 \beta_8 - \beta_2 - \beta_4\label{anomalie_gauss:H,H 13} \\
    \mathcal{A}_{14}^{\left\{ \mathbb{H},\mathbb{H} \right\}}= &\left[ 1 + \beta_2 \right] (2 - \alpha_4 + 3 \alpha_5) - 2 \partial_\mathfrak{c} (\Tilde{g} + \alpha_1) \label{anomalie_gauss:H,H 14}\\
    \mathcal{A}_{15}^{\left\{ \mathbb{H},\mathbb{H} \right\}}=& \left[ 1 + \beta_4 \right] (-2 + \alpha_4 - 3 \alpha_5) + 2 \partial_\mathfrak{c} (\Tilde{g} + \alpha_1) \label{anomalie_gauss:H,H 15}\\
    \mathcal{A}_{16}^{\left\{ \mathbb{H},\mathbb{H} \right\}}=& \partial_\mathfrak{c} \alpha_3 \label{anomalie_gauss:H,H 16}\\
    \mathcal{A}_{17}^{\left\{ \mathbb{H},\mathbb{H} \right\}}=& -6 \left[ \Tilde{g} + \alpha_1 \right] \left[ 1 +\beta_6 \right] + 2 g (\partial_\mathfrak{c}g)(3 +3 \beta_1 - 2 \mathfrak{p} \partial_\mathfrak{p} \beta_1) + g^2 \partial_\mathfrak{c} \beta_1 + 4 g \mathfrak{p} (\partial_\mathfrak{p} g)(\partial_\mathfrak{c} \beta_1) \label{anomalie_gauss:H,H 17}\\
    \mathcal{A}_{18}^{\left\{ \mathbb{H},\mathbb{H} \right\}}=& 12 \Tilde{g} + 10 \alpha_1 + 2 \alpha_6 + 4(\Tilde{g} + \alpha_1) \beta_{10} + 2 (\Tilde{g} + \alpha_6) \beta_2 + 6 \Tilde{g} \beta_9 - 2 \Tilde{g} \partial_\mathfrak{c} \Tilde{g} - 2 g (\partial_\mathfrak{c} g)(5+5 \beta_2 - 2 \mathfrak{p} \partial_\mathfrak{p} \beta_2) \notag\\
    &- \partial_\mathfrak{c} \alpha_2 - g^2 \partial_\mathfrak{c} \beta_2 - 4 g \mathfrak{p} (\partial_\mathfrak{p} g)(\partial_\mathfrak{c} \beta_2)\label{anomalie_gauss:H,H 18} 
\end{align}
\begin{align}
    \mathcal{A}_{19}^{\left\{ \mathbb{H},\mathbb{H} \right\}}= &2 \alpha_1 - 2 \alpha_6 + 6 (\Tilde{g} + \alpha_1) \beta_{11} - 4 ( \Tilde{g} + \alpha_1) \beta_{12} - 2 (\Tilde{g} + \alpha_6) \beta_4 + 2 \Tilde{g} \partial_\mathfrak{c} \Tilde{g} -2 g (\partial_\mathfrak{c} g)(1+\beta_4 + 2 \mathfrak{p} \partial_\mathfrak{p} \beta_4) \notag\\
    &+ \partial_\mathfrak{c} \alpha_2 +g^2 \partial_\mathfrak{c} \beta_4 + 4 g \mathfrak{p} (\partial_\mathfrak{p} g)(\partial_\mathfrak{c} \beta_4) \label{anomalie_gauss:H,H 19}\\
    \mathcal{A}_{20}^{\left\{ \mathbb{H},\mathbb{H} \right\}}= &3 \mathfrak{p} (\Tilde{g} + \alpha_1) + 3 \mathfrak{p} (\Tilde{g} + \alpha_1)\beta_8 - g (\partial_\mathfrak{c}g)\left[ 3 \mathfrak{p} (1+\beta_3) + 2 \mathfrak{p}^2 \partial_\mathfrak{p} \beta_3 \right] + \frac{\mathfrak{p}}{2} g^2 \partial_\mathfrak{c} \beta_3 + 2 \mathfrak{p}^2 g (\partial_\mathfrak{p} g)(\partial_\mathfrak{c} \beta_3).\label{anomalie_gauss:H,H 20} 
\end{align}

\section{Summary of the counterterms in the case of all the constraints being corrected}\label{sec:annexe resume contre-termes}
\numberwithin{equation}{section}
Here $\xi_3(\mathfrak{c},\mathfrak{p})$ is left arbitrary for generality purposes. First, the matter counterterms :

\begin{align}
    &\beta_1 = \beta_2 = \beta_3 = \beta_4 = \beta_5 = \beta_6 = \beta_7 = \beta_8 = 0\\
    &\beta_9 = 1- \partial_\mathfrak{c} (h+\xi_3)\\
    &\beta_{10} = \beta_{11} = \beta_{12} = \partial_\mathfrak{c} (h+\xi_3) -1\\
    &\beta_{13} = \alpha_3 [K, \alpha_9 (\mathfrak{p})]  \partial_\mathfrak{c} (g \partial_\mathfrak{c} g) + \partial_\mathfrak{c} (g \partial_\mathfrak{c} g) -1.
\end{align}

For the Hamiltonian counterterms,

\begin{align}
    &\alpha_1 = g \partial_\mathfrak{c} g - \Tilde{g}\\
    &\alpha_2 = 3 g^2 - \Tilde{g}^2 -2g  (h+\xi_3)\partial_\mathfrak{c} g\\
    &\alpha_3=\frac{1}{\sqrt{\mathfrak{p}}} \left[ K + \int ^p_1 \frac{\alpha_9(u)}{2 \sqrt{u}} du \right]\\
    &\alpha_4=\alpha_5= (\partial_\mathfrak{c}g)^2 + g \partial^2_\mathfrak{c} g -1\\
    &\alpha_6 = 2 g \partial_\mathfrak{c}g - \Tilde{g} - (h+\xi_3) \left[ (\partial_\mathfrak{c} g)^2 + g \partial^2_\mathfrak{c} g \right]\\
    &\alpha_7=\alpha_8=4 g \partial_\mathfrak{c}g \left[ (h+\xi_3) + \mathfrak{p}\partial_\mathfrak{p}(h+\xi_3) \right] - \Tilde{g}^2-2 (h+\xi_3)^2 \left[ (\partial_\mathfrak{c}g)^2 + g \partial^2_\mathfrak{c} g \right] - \left[ g^2 + 4 \mathfrak{p} g \partial_\mathfrak{p} g \right] \partial_\mathfrak{c} (h+\xi_3).\\
    &\alpha_9 = \alpha_9(\mathfrak{p}).
\end{align}

For the gauss counterterms,

\begin{align}
    &\chi_1 = (L-1) \mathfrak{p}\\
    &\chi_2 = L(h+\xi_3)-l.
\end{align}

Finally, the diffeomorphism counterterms,

\begin{equation}
    \xi_1=\xi_2=0,
\end{equation}
    
and 
\begin{equation}
\xi_3=
\begin{cases}
&\xi_3(\mathfrak{c},\mathfrak{p}) \text{  if  } g \partial_\mathfrak{c} g= \Tilde{g}\\
&\mathfrak{c} -h + f(\mathfrak{p}) \text{  if  } g \partial_\mathfrak{c} g \neq \Tilde{g}
\end{cases}
\end{equation}

with $L$, $K$, $f(\mathfrak{p})$ (or $\xi_3(\mathfrak{c},\mathfrak{p})$) and $\alpha_9(\mathfrak{p})$ being degrees of freedom. The only restriction on the correction functions affects $g$ and is given by 

\begin{equation}
    g\partial^2_\mathfrak{c}g+2 \mathfrak{p} \left[ (\partial_\mathfrak{c}g) (\partial_\mathfrak{c}\partial_\mathfrak{p} g)-(\partial_\mathfrak{p} g)(\partial^2_\mathfrak{c} g) \right]=0.
\end{equation}

\end{document}